\pdfoutput=1
\documentclass[aps,twocolumn,prb,groupedaddress]{revtex4}
\usepackage{graphicx}
\usepackage{epstopdf}
\usepackage{amsmath}
\usepackage[T1]{fontenc}

\newcommand{\be}{\begin{equation}}
\newcommand{\ee}{\end{equation}}
\newcommand{\bd}{\begin{displaymath}}
\newcommand{\ed}{\end{displaymath}}
\newcommand{\bk}{\mathbf{k}}
\newcommand{\br}{\mathbf{r}}
\newcommand{\lc}{\lambda_{\rm c}}

\begin{document}
\textheight 25cm 
\sloppy
\title{Polariton states bound to defects in GaAs/AlAs planar microcavities}
\author{Joanna M Zajac}
\email[Electronic address:]{ZajacJM@cardiff.ac.uk}
\author{Wolfgang Langbein}
\affiliation{School of Physics and Astronomy, Cardiff University,
The Parade, Cardiff CF24 3AA, United Kingdom}

\author{Maxime Hugues}
\author{Mark Hopkinson}
\affiliation{Department of Electronic and Electrical Engineering, University of Sheffield. Mappin
Street, Sheffield S1 4JD, UK}

\date{\today}
\begin{abstract}
We report on polariton states bound to defects in planar GaAs/AlAs microcavities grown by molecular beam epitaxy. The
defect types relevant for the spatial polariton dynamics in these structures are cross-hatch misfit dislocations, and
point-like defects extended over several micrometers. We attribute the latter defects to Ga droplets emitted
occasionally by the Ga cell during the growth. These defects, also known as oval defects, result in a dome-like local
modulation of surface, which is translated into the cavity structure and leads to a lateral modulation of the cavity
polariton energy of up to 15\,meV. The resulting spatially localized potential landscape for the in-plane polariton
motion creates a series of bound states. These states were characterized by spectrally resolved transmission imaging in
real and reciprocal space, and reveal the spatial potential created by the defects. Interestingly, the defect states
exhibit long lifetimes in the 10\,ps range, which we attribute to a spatially smooth confinement potential.
\end{abstract}
\maketitle

\section{Introduction}

Semiconductor microcavities have been extensively studied in last 20 years
\cite{KavokinBook07,QCSSSProc09,SkolnickSST98}. Significant attention was given to planar microcavities, which are
Fabry-Perot resonators, with a photon mode confined in growth direction by Bragg mirrors made of pairs of $\lambda$/4
layers of alternating refractive index. The in-plane dispersion of the mode is close to quadratic for in-plane
wavevectors $k$ much smaller than the free-space wavevector, with a curvature described by an effective mass about 4
orders of magnitude smaller than the free electron mass. This implies a significant dispersion within the external
optically accessible wavevector range. When coupled to a quantum well exciton, these microcavities allow to investigate
fundamental condensed-matter phenomena like strong coupling, Bose-Einstein condensation, lasing, parametric
amplification, etc. In many cases, these effects are significantly affected by the disorder in the structures. This
leads to localization of the Bose-Einstein condensate, formation of vortices and can be used as a probe for the
dynamics and excitations the polaritonic superfluid phase \cite{KasprzakNature06,LagoudakisNatPhys08,AmoNatLett09}.

The spatial localization of the cavity mode by artificial in-plane structures was investigated
experimentally and theoretically in Refs.
\onlinecite{KaitouniPRB06,LuganPSSC06,CernaPRB09,CernaPRB10,NardinPRB10}. In these works, circular
mesas with diameters of 3-19\,$\mu$m and heights in the nanometer range were created by lithography
on the cavity layer before growing the top Bragg mirror.

In nominally planar samples it is typically observed that the cavity mode is elastically scattered
on a cross-hatched dislocation pattern \cite{GurioliPRB01,LangbeinPRL02,LangbeinICPS02,
LangbeinJPhys04,LangbeinRNC10}, while the excitonic part tends to exhibit a more isotropic disorder
on the relevant micrometer length scale. In samples grown by MOCVD, additional random disorder is
observed, indicating a fluctuation of the layer thickness in the micrometer spatial range due to
the growth mode influenced by transport of the reactants via the gas-phase, allowing for
non-homogeneous deposition. In MBE instead, the molecular beam guarantees the random deposition of
Ga atoms, and the surface diffusion, which is limited to typically 100\,nm, cannot create a
long-range thickness modulation.

Apart from the cross-hatch dislocation pattern, one typically observes point-like defects (PD) with
a surface density of about $10^4/$cm$^2$, which have a size and shape typical of those ascribed as
oval defects\cite{Herman89MBE}. PDs are better visible in the regions of lower density of
cross-hatches, however were observed in all regions, and for different MC samples which we
investigated until now. They are not related to strain relaxation, but are generally attributed to
particle contamination during growth or Ga source spitting \cite{KawadaJCG93}. These defects
produce rather extended surface modulations, typically without lattice defects. Polariton modes in
MCs have typical extensions of 10-100 micrometers, limited by mirror transmission or residual
disorder. These modes are therefore sensitive to the length-scale of the PDs, making an excellent
probe for structure inhomogeneities in the 1-100 micrometer range. The resonant local transmission
of the MC is modified by the PDs, creating spatially confined cavity modes, typically visible as a
series of spectrally narrow localized modes, as will be shown here. The paper is organized as
follows. In Section\,\ref{sec:setup} the details of the samples and the experiment are given,
followed by a description of the disorder inside these structures in Section\,\ref{sec:disorder},
and a presentation of the results of the optical measurements on localized states in
Section\,\ref{sec:states}. The origin of PDs is discussed in Section\,\ref{sec:PDorigin}.

\section{Sample and Experiment}
\label{sec:setup}
\begin{figure}[t]
\includegraphics[width=\columnwidth]{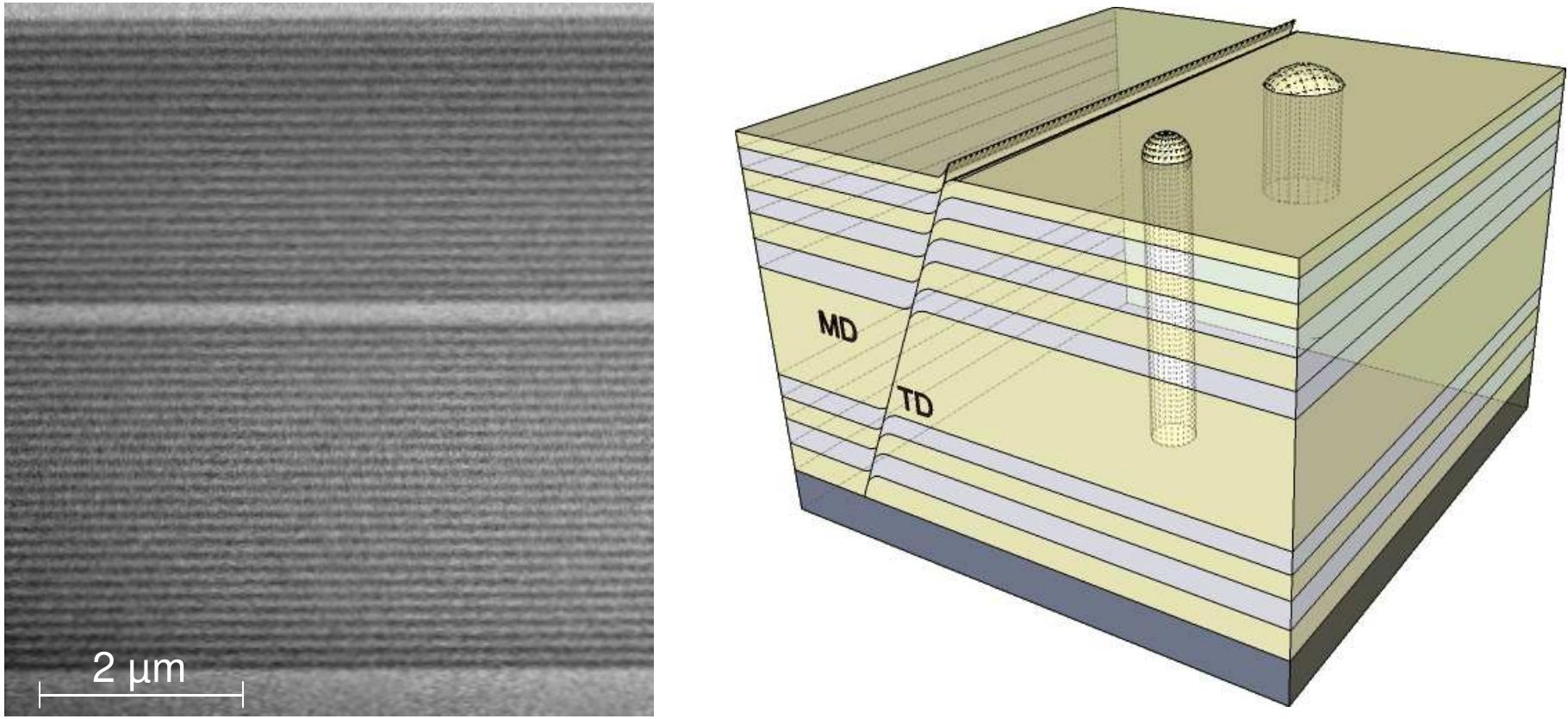} 
\caption{Scanning electron micrograph of the studied microcavity (left), and sketch of the sample
structure (right) showing examples of defects, starting from different layers across the structure.
\label{fig:sample}}
\end{figure}

The sample investigated in this work is a bulk $1\,\lc$ GaAs semiconductor microcavity (MC)
surrounded by AlAs/GaAs distributed Bragg reflectors (DBRs) with 27 (24) $\lc/4$ pairs on the
bottom (top), respectively. Images of a sample are shown on Fig.\,\ref{fig:sample}. At a
temperature of T=80\,K, the cavity mode energy was $E_{\rm C}=1.485$\,eV, red detuned by 23\,meV
from the bulk GaAs exciton in the cavity layer at 1.508\,eV. The structure was grown under
continuous wafer rotation, resulting in a very weak (<1\%) thickness variation from the center to
the edge of the wafer.

\begin{figure}[t]
\includegraphics[width=\columnwidth]{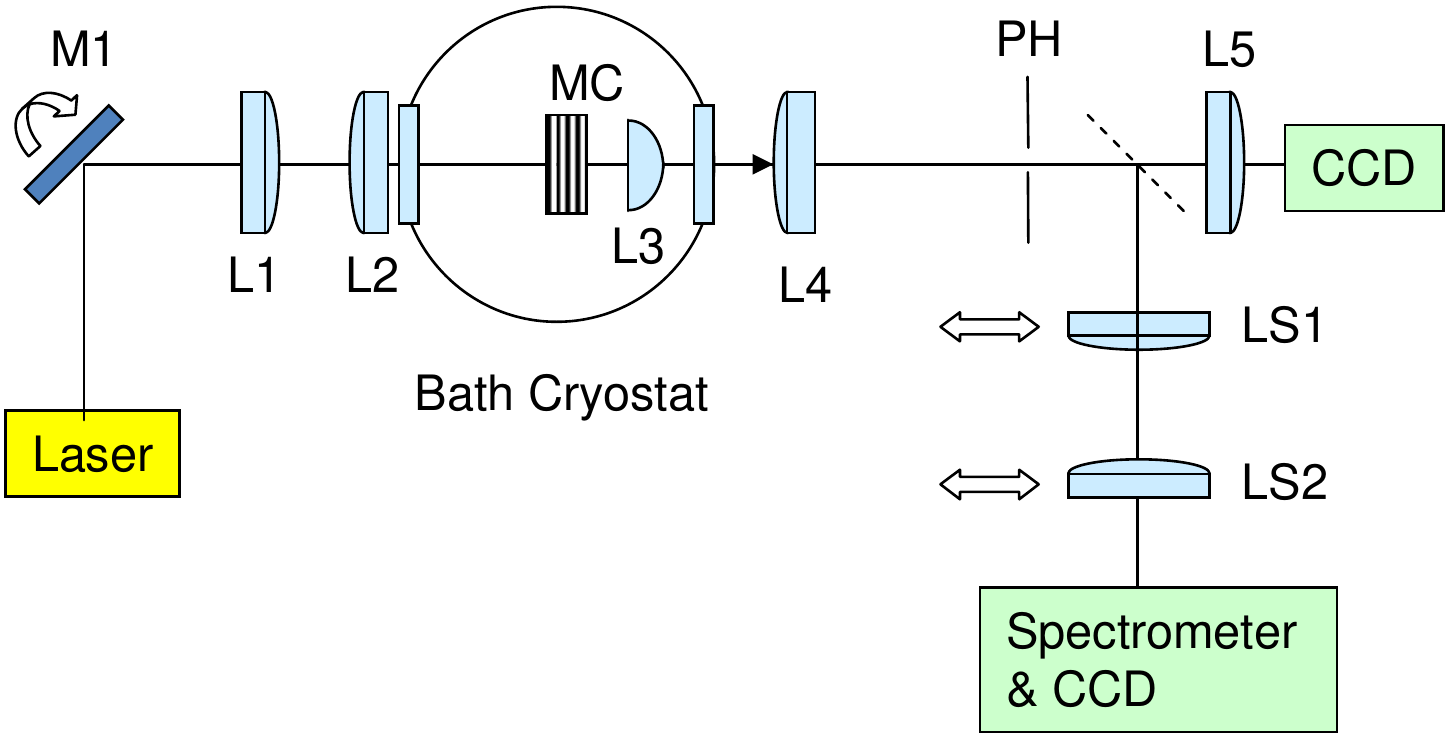} 
\caption{Sketch of the optical setup used to measure the localized polariton states. Both
real-space and reciprocal space imaging was used. L1-L5 Lenses, MC: Microcavity sample LS1,LS2
moving lenses for imaging, for details see text.\label{fig:setup}}
\end{figure}

The optical measurements of the polariton states were conducted at nitrogen temperature (T=80\,K) with the sample mounted strain-free in a bath cryostat. Real and reciprocal space transmission
spectra were taken in the low-intensity regime. A sketch of the setup is shown on
Fig.\,\ref{fig:setup}.

Two different optical excitation configurations were used. Firstly, a single-mode external cavity diode laser (Sacher
Lynx) with 5\,MHz linewidth was employed to excite the polaritons at normal incidence ($\bk=0$) over a large size in
real space ($\sim 1\,$mm) from the substrate side. This excitation allowed to selectively image a specific polariton
energy with a spectrally integrating detector. Secondly, a mode-locked Ti:Sapphire laser (Coherent Mira) was used
providing 100\,fs pulses at 76\,MHz repetition rate. The large spectral width of approximately 20\,meV allows to excite
polariton bound states and the continuum dispersion simultaneously. The excitation was spatially focussed to a spot of
about $3\,\mu$m diameter using a lens of 0.15 numerical aperture, corresponding to an excitation wavevector range of
$|\bk|<1.1/\mu$m. The emission was collected from the epi-side by an aspheric lens of 0.5\,NA (L3 in
Fig.\,\ref{fig:setup}) with a wavevector range of $|\bk|<4/\mu$m, providing a spatial resolution of about 1\,$\mu$m.
The emission was detected spectrally integrated in the real or reciprocal space using video CCD cameras. For spectrally
resolved detection, the emission was imaged in real or reciprocal space into the input slit of an imaging spectrometer
with 12\,$\mu$eV spectral resolution. Two-dimensional images were acquired by translating the image across the
spectrometer slit using lateral movements of two lenses (LS1 and LS2 in Fig.\,\ref{fig:setup}). Scans were made with
$0.5\,\mu$m (real space), or $0.05\,\mu$m$^{-1}$ (reciprocal space) steps.

\section{Disorder variation versus growth temperature}
\label{sec:disorder}
\begin{figure}[t]
\includegraphics*[width=8cm]{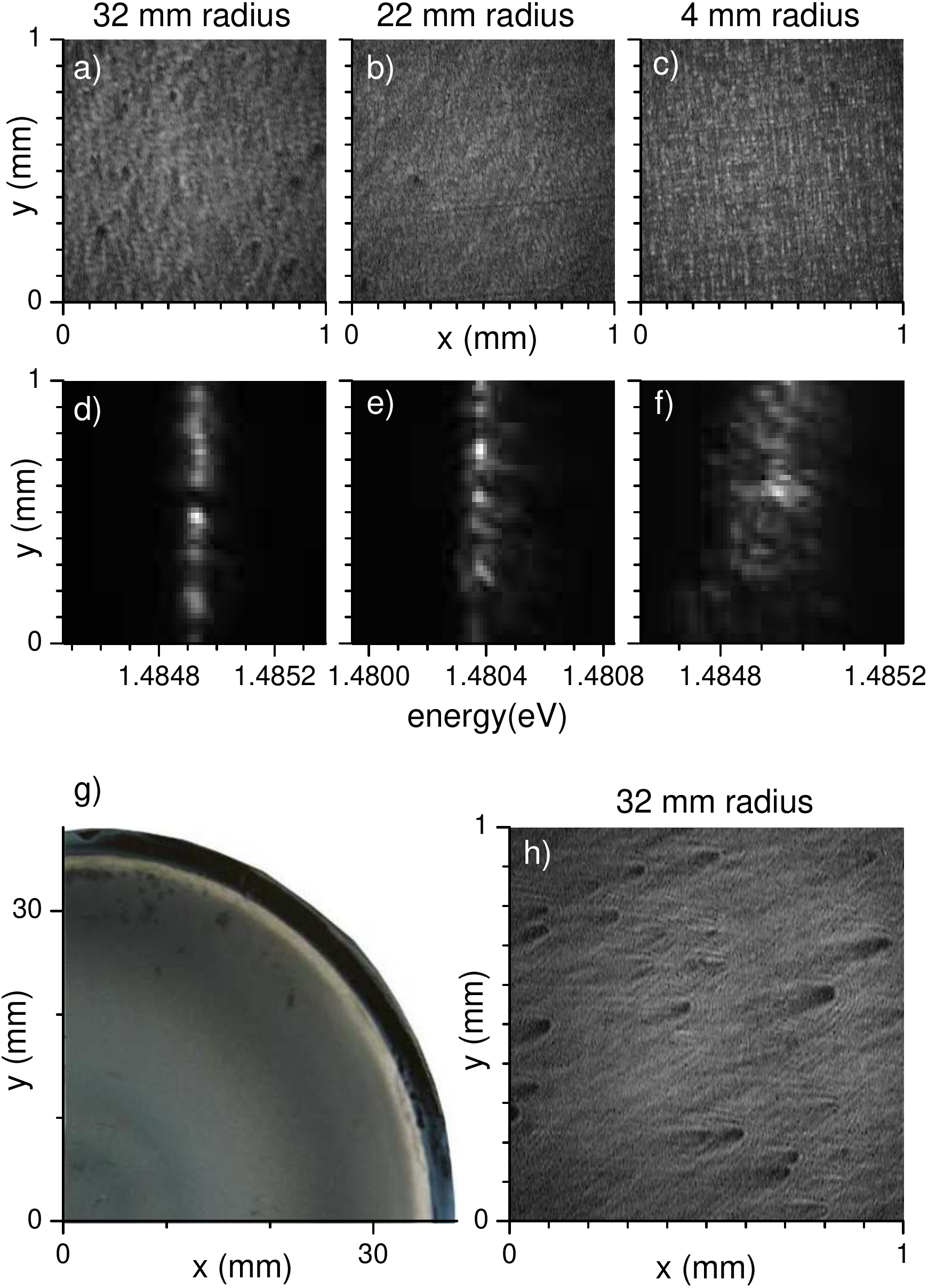}
\caption{Disorder variation across the 3\,inch wafer. Transmitted intensity for excitation at
$\bk=0$ with 100\,fs pulses resonant to the cavity mode for samples from different radial positions
on the wafer as indicated. a),b),c) spatially resolved, spectrally integrated. d),e),f) spectrally
resolved as function of $y$ at a given $x$ position. Linear gray scale from zero (black) to white.
g) Image of the back side of the wafer. h) as a), but for excitation at $\bk\approx (-0.3,0)/\mu$m.
\label{fig:disorder_radial}}
\end{figure}

During the growth of the investigated structure, the temperature was ramped up to $715\,^\circ $C for the AlAs layers,
and down to $660^\circ $C for the Bragg GaAs layers, and to $630^\circ $C for the GaAs cavity layer. For comparison, in
Ref.\,\onlinecite{OesterlePSSB05} the sample was grown at a constant temperature of $620-630^\circ $C. A radial
non-uniformity of the back-surface roughness due to As evaporation and resulting Ga-droplet formation was observed (see
Fig.\,\ref{fig:disorder_radial}g). This indicates that the growth temperature was non-uniform across the wafer, an
effect which is significant due to the usage of an undoped and back-side polished 3-inch wafer, leading to a weaker
radiative coupling to the surrounding compared to doped wafers, increasing the infrared absorption, and unpolished
wafers, decreasing the reflection and avoiding radiation trapping by total-internal reflection. The cavity resonance
energy was varying by only 0.5\,\% over the wafer, indicating an exceptional flux homogeneity of the Ga and Al cells.
The polariton disorder instead was found to vary significantly as function of radial position $R$ on the wafer, as
shown in Fig.\,\ref{fig:disorder_radial}a)-f) where transmission images and spectra are shown at different $R$. Close
to the center ($R=4$\,mm), a cross-hatch pattern is visible and the polariton states excited at $k=0$ show an
inhomogeneous broadening of $\sim 100\,\mu$eV. This cross-hatch disorder decreases with increasing radius, and is not
discernible at $R=32$\,mm close to the edge of the wafer. The cross-hatch pattern is due to the formation of a
misfit-dislocation pattern found in (001) oriented strained cubic semiconductor thin film systems\cite{AndrewsJAP02}.
It is formed by strain relaxation via slides along the \{111\} slip planes, which form lines along $[110]$ and
$[1\bar{1}0]$ at their intersection with the film surface. Such a pattern is only observed for weakly strained films
($\epsilon<2\%$), while for higher strains Stranski-Krastanov or Volmer-Weber growth is found. Cross-hatch patterns are
typically reported for a lattice mismatch of the order of 1\%, leading to a large density of slip lines and a
significant surface modulation of several nanometers. The lattice mismatch of the GaAs/AlAs Bragg mirrors at room
temperature\cite{AdachiBook88} is 0.14\%. Using a mismatch of 0.07\% of a Bragg period, we find a critical thickness
for strain relaxation\cite{MatthewsJCG74} of about 2\,$\mu$m. The Bragg mirrors have a total thickness of 6\,$\mu$m,
exceeding the critical thickness, and misfit dislocations form. However, the mismatch decreases with temperature
\cite{EttenbergJAP70,AdachiBook88}, providing a lattice matched system at about 900\,$^\circ$C. We can therefore expect
that the growth temperature influences the cross-hatch formation.

In the wafer region with low cross-hatch disorder ($R=32$\,mm) the polariton linewidths are within
30\,$\mu$eV over extended regions. This exceptional spatial homogeneity is confirmed by the
propagation of polaritons over hundreds of micrometers visible in Fig.\,\ref{fig:disorder_radial}\,h),
where the interference pattern of freely propagating polaritons with the ones scattered by the PDs is
observed for oblique excitation a few degrees from normal corresponding to $\bk\approx (-0.3,0)/\mu$m.
PDs are visible in all sample regions, and form natural defects which we investigate further in the
subsequent sections.

\section{Polariton states bound to point-like Defects}
\label{sec:states}
The spatial distribution of polariton energies can be visualized by resonant excitation with a
spectrally narrow source. We used a single-mode external cavity diode laser (Sacher Lynx) with
5\,MHz linewidth to excite the polaritons at normal incidence ($\bk=0$) over a large size in real
space ($\sim 1\,$mm) from the substrate side. The emission from the epi side was imaged onto a CCD.
By tuning the excitation photon energy below the band edge of the extended cavity polaritons,
individual localized states can be excited resonantly and appear as bright spots. By scanning the
photon energy, the localized defect states within the excited area are sequentially addressed
according to their eigenenergy.
\begin{figure}[t]
\includegraphics[width=8cm]{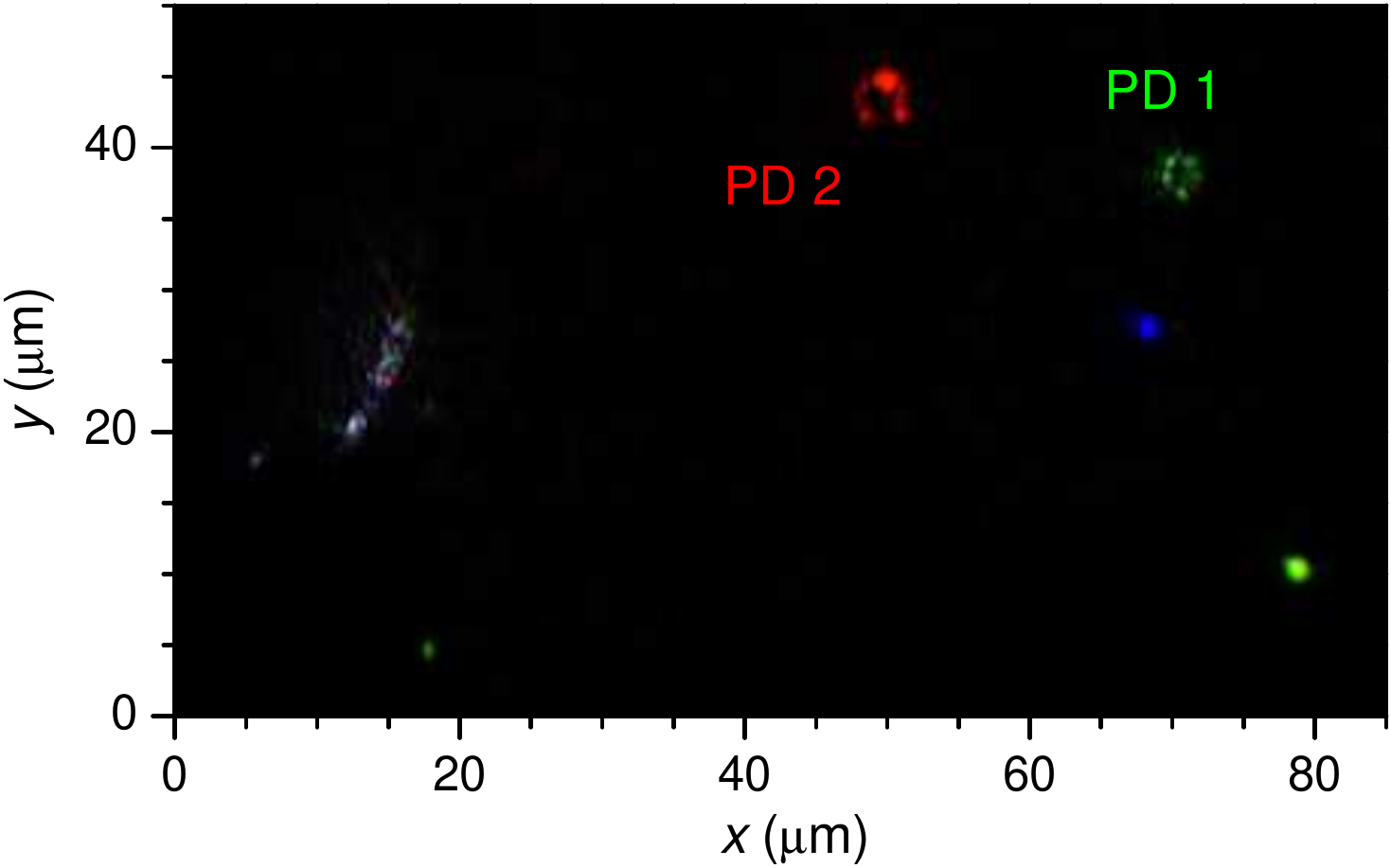}
\caption{Spatially resolved color coded transmission intensity for photon energies of 1482.1\,meV
(blue channel), 1482\,meV (magenta channel), and 1473.8\,meV (yellow channel), see movie in supplementary
materials where relevant energy tuning range is 1482.1 - 1473.8 eV. Two localized states are labeled PD\,1 and PD\,2 for later reference. \label{fig:cwimage}}
\end{figure}

An example is shown in Fig.\,\ref{fig:cwimage}, where images corresponding to three different excitation
photon energies are overlayed into a color image. The different colors of the localized states show their
different eigenenergies $\hbar\omega_n$. Different shapes of the localized states are also visible, and we
find an average distance of a few $10\,\mu$m of the PDs with localized states separated by more than 1\,meV
from the band edge. To gain more detailed information about the set of states bound to an individual PD, we
change excitation and detection setup to the pulsed 100\,fs source centred $\sim 10$\,meV below the
polariton band-edge, spatially focussed onto the PD. The emission was measured using spectrally resolved
imaging in either real or reciprocal space, as discussed in section \ref{sec:setup}.

The resulting data for the polariton states bound to PD\,1 are shown in Fig.\,\ref{fig:spectralimaging3}. On the left
the directionally and spectrally resolved intensity $I(k_x=0,k_y,\omega)$ is given on a logarithmic scale showing bound
resonances and the continuum. A series of $n_{1}=15$ discrete resonances is discernable, at frequencies $\omega_n$,
$n=1..n_1$. The intensity distributions at these resonances in real and reciprocal space, $I(x,y,\omega_n)$ and
$I(k_x,k_y,\omega_n)$, respectively, are proportional to the probability distributions $\left|\Psi_n\right|^2$ of the
corresponding state wavefunctions $\Psi_n$ (assuming that the Bragg mirror reflectivity is independent of $\bk$ and all
relevant emission directions $\bk$ are imaged). For convenience we show the localization energy
$\Delta_n=\hbar\omega_n-\hbar\omega_{\rm c}$ of the states, where $\hbar\omega_{\rm c}$ is the band edge given by the
minimum of the parabolic dispersion of the free polaritons.

\includegraphics[width=8.4cm]{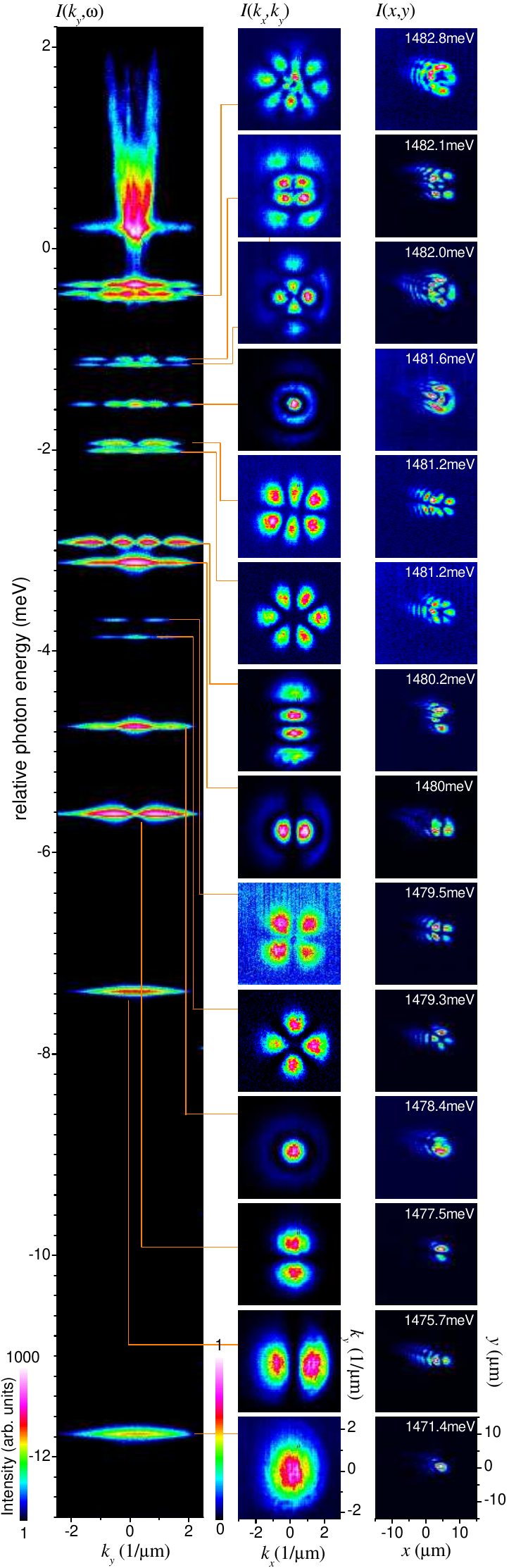}
\begin{figure}
\caption{ Spectral imaging of the emission from polariton states close to PD\,1 of
Fig.\,\ref{fig:cwimage} in real and reciprocal space. Left: intensity $I(k_x=0,k_y,\omega)$ on a
logarithmic color scale as indicated. The energy $\hbar\omega$ is shown relative to the polariton
band edge at $\hbar\omega_{\rm c}=1.4833$\,eV. The intensity of the individual states is given in
reciprocal space $I(k_x,k_y,\omega_n)$ in the middle column and in real space $I(x,y,\omega_n)$ in
the right column. They are shown using a linear color scale and are individually normalized to
their maximum. The resonance energies $\hbar\omega_n$ are given, and the connection lines indicate
the corresponding state in $I(k_x=0,k_y,\omega)$. The tails visible in $I(x,y,\omega_n)$ are due to
imaging abberations relevant at large $\bk$.} \label{fig:spectralimaging3}
\end{figure}

The ground state $\Psi_1$ is observed at $\Delta_1=-11.9$\,meV. Some resonances are observed even above the continuum,
indicating the presence of a potential barrier between the defect region and the surrounding continuum. These states
could lead to resonant scattering similar to Feshbach resonances observed in atomic physics. The shapes of the
wavefunctions reveal a nearly cylindrical symmetry of the effective confinement potential $V_1(x,y)$ created by PD\,1.
The first excited state $\Psi_2$ is found at $\Delta_2=-7.6$\,meV, and shows a $p_x$-like symmetry, with a node at
$x=0$. $\Psi_3$ has $\Delta_3=-5.8$\,meV and a $p_y$-like symmetry with a node at $y=0$. The energy splitting of
$\Psi_{2,3}$ shows a breaking of cylindrical symmetry of $V_1(x,y)$. The next 3 states correspond to d-states,
$\Psi_{4}$ having zero angular momentum and 2 nodes in radial direction, and $\Psi_{5,6}$ being superpositions of
angular momentum 2. The non-degeneracy of the different n=2 states shows the non-parabolicity of the confinement. The
higher states can be classified in a similar way with increasing number of nodes in the wavefunction.

\begin{figure}[t]
\includegraphics[width=\columnwidth]{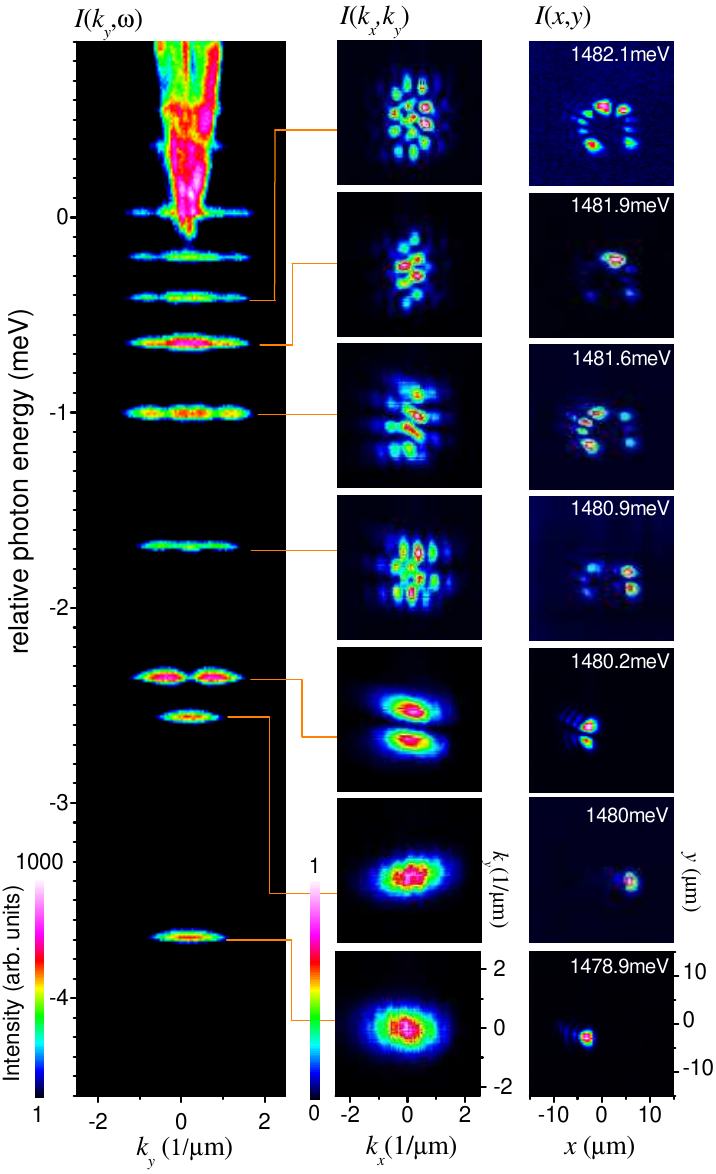}
\caption{As figure \ref{fig:spectralimaging3}, but for PD\,2 of Fig.\,\ref{fig:cwimage} and
$\hbar\omega_{\rm c}=1.4826$\,eV.} \label{fig:spectralimaging1}
\end{figure}

PD\,2 instead is not cylindrically symmetric as shown in Fig.\,\ref{fig:spectralimaging1}. The two
lowest states $\Psi_{1,2}$ are separately localized to the left and the right of the defect center
with a s-like wavefunction, indicating two separate minima of the potential. The left minimum
confines also a p-like excited state $\Psi_{3}$, while the corresponding excited state on the right
$\Psi_{4}$ is already extended along a horseshoe-shaped region, coupling to a d-like state of the
left minimum. The next state $\Psi_{5}$ is a mixture of the left f-like state with the right d-like
state.  $\Psi_{6}$ is mostly localized on the tip of the horseshoe, possibly due to a local
potential maximum close to the state energy, which is supported by the relatively small wavevector
spread in $k_x$. A total of $n_{3}=11$ states are visible.

\begin{figure}[t]
\includegraphics[width=\columnwidth]{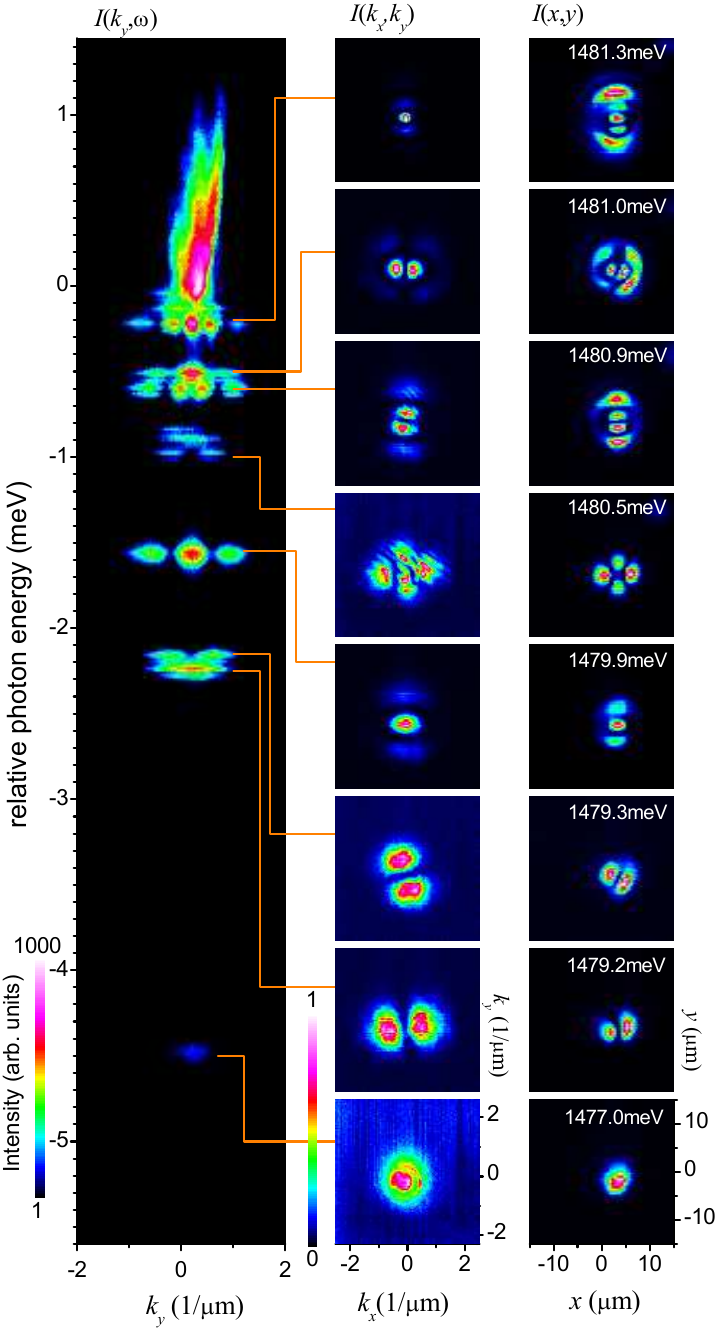}
\caption{As figure \ref{fig:spectralimaging3} but for PD\,3, (not shown on Fig.\,\ref{fig:cwimage}), and
$\hbar\omega_{\rm c}=1.4815$\,eV.} \label{fig:spectralimaging2}
\end{figure}

The last reported defect, PD\,3 (see in Fig.\,\ref{fig:spectralimaging2}, not shown in Fig.\,\ref{fig:cwimage}) is
similar to PD\,1 in being approximately circular symmetric, but about a factor of 3 shallower. The ground state
$\Psi_{1}$ is s-like, and shows a larger extension in real space and a smaller in reciprocal space than the ground
state of PD1, as expected from the weaker confinement. A total of $n_{3}=12$ states are visible.

\subsection{Coherent wavepacket dynamics}
The two p-like excited states $\Psi_{2,3}$ of PD\,3 are nearly degenerate ($\approx100\,\mu$eV
separation). Due to the coherent excitation of the states and their finite spectral width, the
emitted field is a coherent superposition of the two wavefunctions $a\Psi_2 + b\Psi_3$, with
amplitudes determined by the excitation. To compare the measurements with the theoretically
expected result, we assume harmonic confinement and p-like wavefunctions
$\Psi_{2,3}\propto(k_{x},k_{y})\exp\left(-\frac{k_{x}^2+k_{y}^2}{a^2}\right)$ with frequencies
$\omega_{2,3}=\omega_0\pm\Delta_{\rm p}$ and linewidths $\gamma_{\rm p}$ which are a sum of the
state linewidth and spectrometer resolution. The measured wave-function can be modeled as a
coherent superposition of the two near-resonant states with a relative phase $\phi$ and amplitude
$\eta$ given by the excitation conditions. This superposition can be written as

\begin{eqnarray}
\Psi(\omega)&\propto&\big(\frac{k_{x}}{\omega-(\omega_0-\Delta_{\rm p})+i\gamma_{\rm p}}+
\nonumber \\
&&\eta e^{i\phi}\frac{k_{y}}{\omega-(\omega_0+\Delta_{\rm p})+i\gamma_{\rm p}}
\big)e^{-\frac{k_{x}^2+k_{y}^2}{a^2}}
\end{eqnarray}

Similarly, for the d-like $l=2$ states
$\Psi_{5,6}=(k_{x}k_{y},k_{x}^2-k_{y}^2)\exp\left(-\frac{k_{x}^2+k_{y}^2}{a^2}\right)$ with frequencies
$\omega_{5,6}=\omega_0\pm\Delta_d$ and linewidth $\gamma_{\rm d}$,

\begin{eqnarray}
\Psi(\omega)&\propto&\big(\frac{k_{x}k_{y}}{\omega-(\omega_0-\Delta_{\rm d})+i\gamma_{\rm d}}+
\nonumber \\
&&\eta e^{i\phi}\frac{k_{x}^{2}-k_{y}^{2}}{\omega-(\omega_0+\Delta_{\rm d})+i\gamma_{\rm
d}}\big)e^{-\frac{k_{x}^2+k_{y}^2}{a^2}}
\end{eqnarray}

\begin{figure}
\includegraphics[width=\columnwidth]{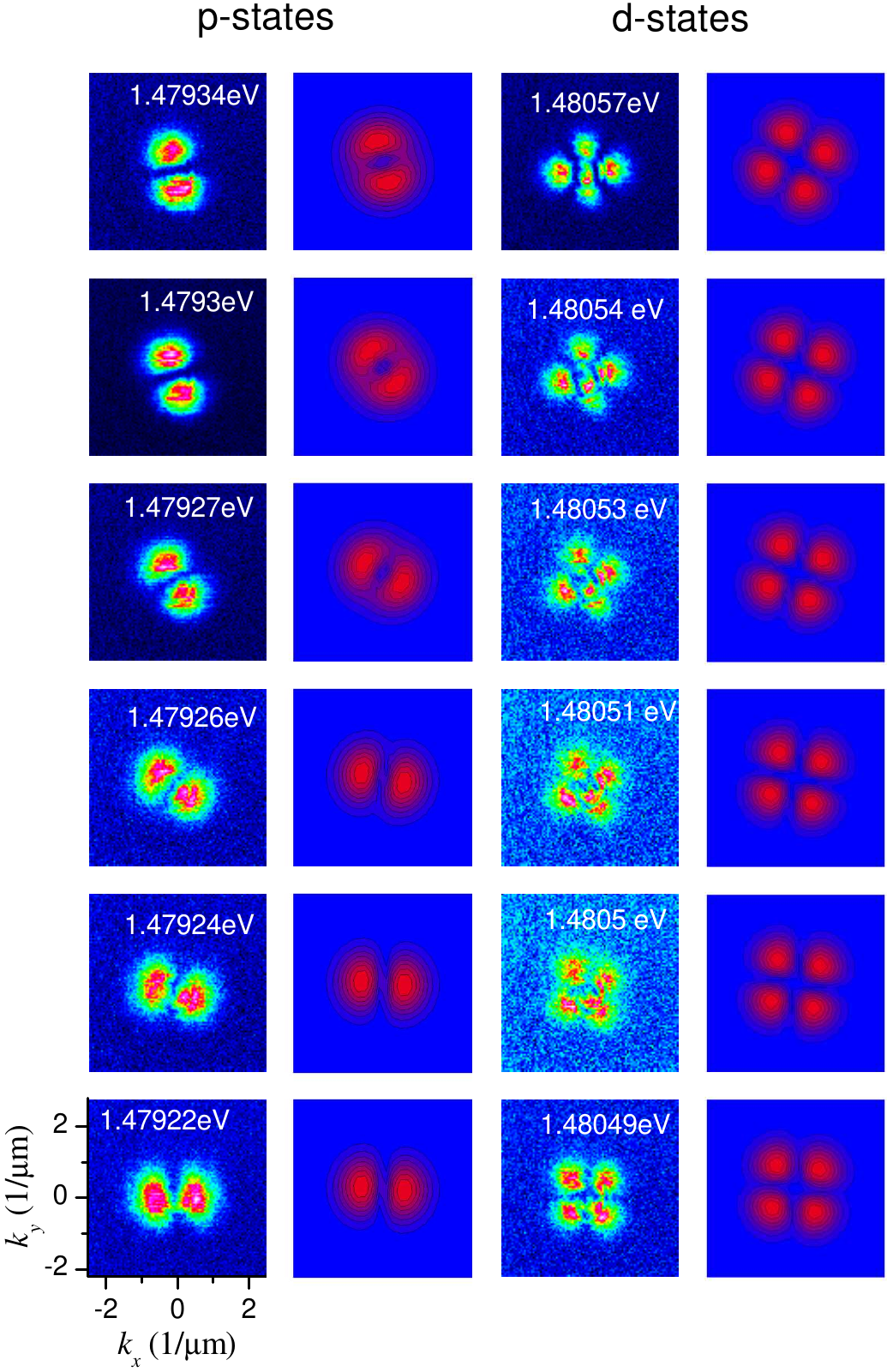}
\caption{Measured intensities $I(\bk,\omega)$ (left columns) and simulated $|\Psi(\omega)|^2$ for
energies close to the p-states (left) and d-states (right). The energy is increasing from lower to
upper panels as labeled. The parameters used were $\hbar\omega_0=1.479205\,$eV,
$\hbar\Delta=90\,\mu$eV, $\hbar\gamma_{1}=58\,\mu$eV, $\hbar\gamma_{2}=73\,\mu$eV, $\eta=1$,
$\phi=0$ for the p-states, and $\hbar\omega_0=1.48047\,$eV, $\hbar\Delta=89\,\mu$eV,
$\hbar\gamma_{1}=37\,\mu$eV, $\hbar\gamma_{2}=50\,\mu$eV,$\eta=1$, $\phi=0$   for the d-state
simulation. \label{fig:simulation_pd}}
\end{figure}

Simulations were done using parameters values as indicated in Fig.\,\ref{fig:simulation_pd}, and give a qualitative
agreement with measurements. A movie over the detuning is available in the online material.

\begin{figure}
\includegraphics[width=\columnwidth]{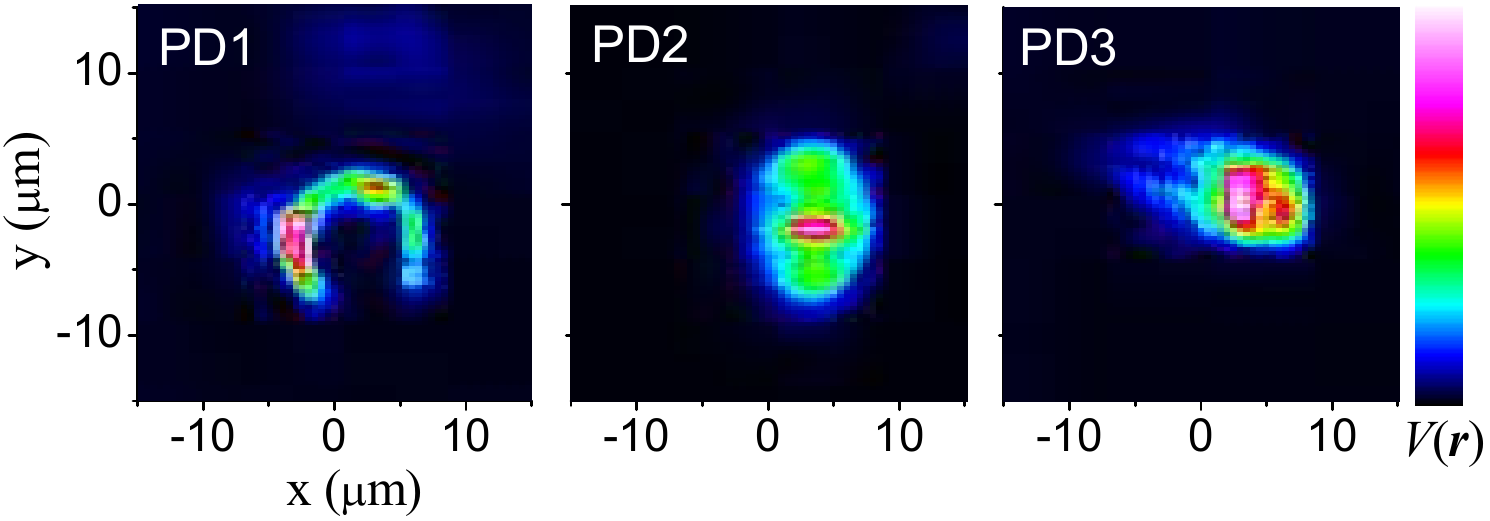}
\caption{Confinement potential $V_m(\br)$ estimated from the bound density of states $D_m(\br)$ for the three PDs, on a
color scale as given from 0 (black) to -2.3\,meV. \label{fig:potentials}}
\end{figure}

\subsection{Confining potential}
The observed localized polariton states can related to an effective confinement potential $V_m(\br)$  for the in-plane
polariton motion. We can estimate $V_m(\br)$ using the spectrally integrated density of states $D_m(\br)$ created by
$V_m(\br)$ below the continuum edge in the following way. On the one hand, this density can be calculated from the
spatially resolved bound density of states
\be D_m(\br)=\sum_{n=1}^{n_{\rm m}} |\Psi_n(\br)|^2 \ee
where the bound state probability densities $|\Psi_n(\br)|^2$ are taken as the normalized measured
intensity
\be |\Psi_n(\br)|^2=\frac{I(\br,\omega_n)}{\int I(\br,\omega_n) d\br^{2}}\ee
This expression assumes that the emission intensity is proportional to the polariton wavefunction, which is valid for a
cavity lifetime which is constant for the in-plane wavevector components of the bound states. This is a reasonable
approximation for the data shown here.

On the other hand, $D_m(\br)$ is given by the integral of the free density of states from zero kinetic energy at the
potential floor to the continuum when neglecting the spatial variation of the confinement potential, {\em i.e.} in the
limit of small level splitting compared to the confinement potential. In two dimensions the density of states is
constant and given by $D_{2D}=m/(2\pi\hbar^2)$, such that the integral is simply given by $-V_m(\br)D_{\rm 2D}$, and we
find $V_m(\br)=-D(\br)/D_{\rm 2D}$. We use the effective mass of the polaritons from the measured dispersion
$m=2\cdot10^{-5}\,m_{\rm e}$, where $m_{\rm e}$ is the free electron mass. The resulting confinement potentials for the
three investigated PDs are shown in Fig.\,\ref{fig:potentials}. The symmetry of the potentials reflect the symmetry of
the localized states. Comparing with the energies of the confined states, $V_m(\br)$ seems generally a factor of 2 to 3
to small. Errors in the scaling of $V_m(\br)$ are related to the length calibration of the imaging, which we estimate
to have an accuracy of $10\%$, yielding a $20\%$ error in $V_m(\br)$, and to the effective mass $m$ which we estimate
to have an error of $10\%$, resulting in a total of 30\% systematic error of the scaling. The finite amount of levels
in the potential leads mostly to residual spatial oscillations, while the average potential is reproduced to a relative
error of $1/n_{\rm m}$, so about 10-20\% in our case. A direct comparison of the polariton states in the confining
potential $V_m(\br)$ could be used to determine the underlying potential in a more quantitative fashion.

\section{Origin of the point-like defects}
\label{sec:PDorigin}
The formation of the PDs could be due to a variety of different physical processes. Among them,
there are threading dislocations(TD), either propagating from the substrate or created in the
epilayers due to surface defects. Usually by growing multilayer structures threading dislocations
from the substrate are suppressed as they bend on GaAs/AlAs interfaces and propagate parallel to
the $\langle$110$\rangle$ directions suppressing the dislocation density even to 5\% of its
original value \cite{ShinoharaJAP85}. The investigated sample was grown on a wafer with a TD
density of about $2\cdot10^3/$\,cm$^2$ specified by manufacturer, which is comparable to the
observed defect density.

\begin{figure}
\includegraphics[width=\columnwidth]{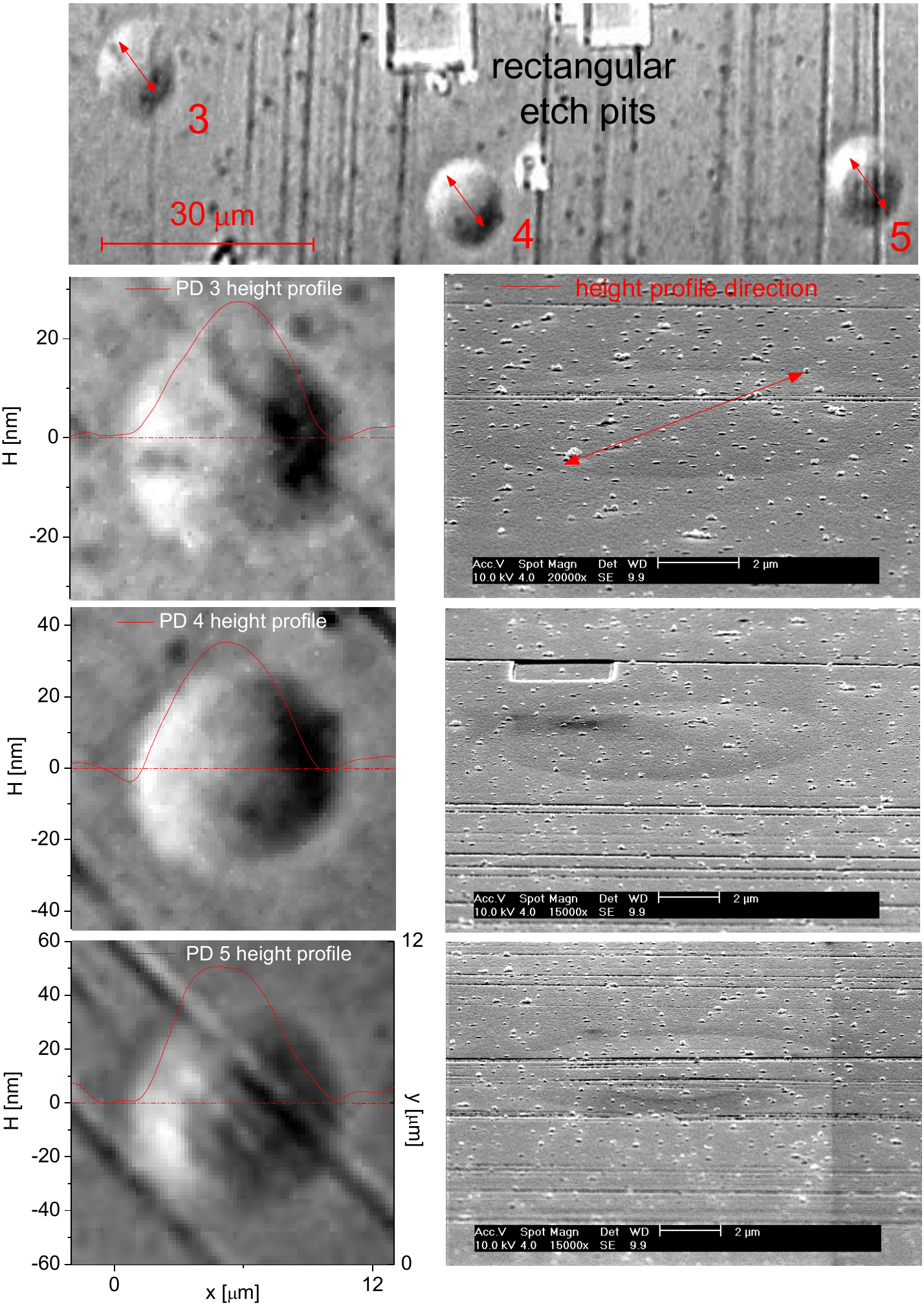}
\caption{Top: Sample surface after KOH etching imaged with DIC. Round defects corresponding to PDs
are labeled. Below on the left column,individual height profiles of PDs obtained using DIC with the
corresponding DIC images. In the right column, corresponding FESEM images of the PDs.}
\label{fig:DIC}
\end{figure}

In order to investigate if the observed PD are related to threading dislocations, we used an etching procedure
\cite{GrabmaierPSSB69,BacherJElChemSoc95}, which requires less equipment and sample preparation as compared to the
alternative method of transmission electron microscopy. It also allows to investigate large surface areas, which is
necessary considering the small defect density. Since the sample is not destroyed in the process, it is possible to
spatially correlate the etched pits corresponding to the treading dislocations with the PDs. We used the procedure
described in Ref.\,\onlinecite{GrabmaierPSSB69} for GaAs etching of \{100\} crystal facets. The sample was kept in KOH
etch at 360\,$^o$C for 2 minutes. The resulting etch-pits on the sample surface were observed in a reflective DIC
(Differential interference contrast) optical microscope. We found an etch-pit density of about $10^{3}$/\,cm$^2$,
consistent with the manufacturer specification. An overview of several defects obtained using DIC microscopy is shown
in figure Fig.\,\ref{fig:DIC}. Each rectangular etch pit marks a threading dislocations in its center. The lines are
scratches on the sample surface and/or misfit dislocations.

Subsequently the low temperature transmission measurements were repeated together with reflection microscopy. The
positions of the PDs were found to be not correlated to the etch-pits. Instead, the PDs coincide with the round to oval
structures of 7-10 micrometer diameter on the sample surface, as observed in Fig.\,\ref{fig:DIC}. These defects are
thus the origin of the localization potentials for the polaritons.

From the DIC images, we have quantitatively extracted the surface profile of the PDs. They show a
typical surface height modulation of several tens of nanometers, being approximately parabolic with
a small depression about 10\% of the height below the original surface at the edge, and a width of
a few $\mu$m. The energy barrier to the continuum observed in the localized states (see Section
\ref{sec:states}) is likely to be related to this edge depression.

Also atomic force microscopy images of the surface were obtained, but due to the large extension and small height of
the PDs no reliable results of the large scale topology could be extracted.

Using field emission scanning electron microscopy(FESEM), the structures were imaged under grazing angle of incidence
(70$^{\circ}$ to the surface normal) to enhance the contrast for the flat surface topology, and results are shown in
Fig.\,\ref{fig:DIC}. The shape is consistent with the DIC images. A ridge shaped structure close to the center of the
dome is also visible.

Growth related defects were categorized in Ref.\,\onlinecite{KawadaJCG93}, showing some defects which qualitatively
resemble the ones discussed here. We observed defects with different shapes, with the most common being nearly round
defects with a weak ridge in the center, as shown on Fig.\,\ref{fig:DIC}. According to Ref.\,\onlinecite{KawadaJCG93}
round defects with nucleus originate from Ga oxide or spitting, while obscure ovals arise from Ga droplets, oxides or
particle contaminants. We attribute the observed defects to Ga nano-droplets emitted by the the Ga cell during growth.
Once deposited on the surface, the Ga is converted to GaAs, forming a hill with a depression in the center due to the
growth kinetics. During the subsequent growth, lateral Ga diffusion leads to an extension of the hill while its height
is reduced. Interestingly, we observe nearly round defects in the present sample, while on similar samples grown
throughout at a lower temperature of $630^\circ$C similar defects are observed, which are elliptical with an about 1:3
ratio of extension in the $[110]$ to $[1\bar{1}0]$ directions. We attribute this observation to the
temperature-dependent anisotropic Ga mobility on the surface.

Clearly the impact of such a structure on the polaritons depends on the position of the Ga droplet deposition within
the MC structure in growth direction. This information is not accessible in the characterization techniques reported
here, which are all measuring the surface topology of the structure. To investigate this further, we used combined
focussed ion beam (FIB) - scanning electron beam (SEM) to cut with FIB into the structure and measure the exposed
side-wall by SEM \cite{Giannuzzi05FIB}. By sequentially slicing the structure, a volume information on the defect is
obtained. The results of this investigation including a study on the their influence of the droplet positions relative
to the cavity layer on polariton confinement will be reported in a forthcoming work.

\section{Summary}
In summary, we have identified point-like defects in planar microcavities which lead to series of spectrally sharp
localized polariton states. The localized polariton states were characterized in real and reciprocal space and the
underlying confinement potential has been reconstructed from these measurements. The symmetry of the states can be very
close to cylindrical. The origin of the defects is attributed to spitting of the Ga cell during MBE growth, depositing
Ga droplets of a size of the order of 100\,nm onto the sample surface during growth. Due to the large thickness of
microcavity structures, the surface density of these droplets is enhanced, and the spatially extended polariton states
are exceptionally sensitive to the resulting extended structural variations. The induced confinement potential can be
rather smooth, allowing for confined states of small linewidth since the scattering into the lossy microcavity modes at
high wavevectors is suppressed.
\acknowledgments We thank A.\,Drevisi and I.\,Fallis for assistance with the KOH etching. This work was supported by
the EPSRC UK Research Council (grant n. EP/F027958/1). Author contributions: WL designed the samples, MH and MH grew
the samples at the EPSRC National Centre for III-V Technologies, Sheffield, UK, and gave feedback on sample design, JZ
performed the structural measurements, JZ and WL performed the optical measurements, analyzed the data and prepared the
manuscript.
%

\begin{thebibliography}{27}
\expandafter\ifx\csname natexlab\endcsname\relax\def\natexlab#1{#1}\fi \expandafter\ifx\csname
bibnamefont\endcsname\relax
  \def\bibnamefont#1{#1}\fi
\expandafter\ifx\csname bibfnamefont\endcsname\relax
  \def\bibfnamefont#1{#1}\fi
\expandafter\ifx\csname citenamefont\endcsname\relax
  \def\citenamefont#1{#1}\fi
\expandafter\ifx\csname url\endcsname\relax
  \def\url#1{\texttt{#1}}\fi
\expandafter\ifx\csname urlprefix\endcsname\relax\def\urlprefix{URL }\fi \providecommand{\bibinfo}[2]{#2}
\providecommand{\eprint}[2][]{\url{#2}}

\bibitem[{\citenamefont{Kavokin et~al.}(2007)\citenamefont{Kavokin, Baumberg,
  Malpuech, and Laussy}}]{KavokinBook07}
\bibinfo{author}{\bibfnamefont{A.}~\bibnamefont{Kavokin}},
  \bibinfo{author}{\bibfnamefont{J.~J.} \bibnamefont{Baumberg}},
  \bibinfo{author}{\bibfnamefont{G.}~\bibnamefont{Malpuech}}, \bibnamefont{and}
  \bibinfo{author}{\bibfnamefont{F.~P.} \bibnamefont{Laussy}},
  \emph{\bibinfo{title}{Microcavities}} (\bibinfo{publisher}{OUP Oxford},
  \bibinfo{year}{2007}).

\bibitem[{\citenamefont{B.~Deveaud-Plédran and
  Schwendimann}(2009)}]{QCSSSProc09}
\bibinfo{editor}{\bibfnamefont{A.~Q.} \bibnamefont{B.~Deveaud-Plédran}}
  \bibnamefont{and}
  \bibinfo{editor}{\bibfnamefont{P.}~\bibnamefont{Schwendimann}}, eds.,
  \emph{\bibinfo{title}{Quantum Coherence in Solid State Systems}}, vol.
  \bibinfo{volume}{171} of \emph{\bibinfo{series}{International School of
  Physics Enrico Fermi}} (\bibinfo{publisher}{IOS Press},
  \bibinfo{year}{2009}).

\bibitem[{\citenamefont{Skolnick et~al.}(1998)\citenamefont{Skolnick, Fisher,
  and Whittaker}}]{SkolnickSST98}
\bibinfo{author}{\bibfnamefont{M.~S.} \bibnamefont{Skolnick}},
  \bibinfo{author}{\bibfnamefont{T.~A.} \bibnamefont{Fisher}},
  \bibnamefont{and} \bibinfo{author}{\bibfnamefont{D.~M.}
  \bibnamefont{Whittaker}}, \bibinfo{journal}{Semicond. Sci. Technol.}
  \textbf{\bibinfo{volume}{13}}, \bibinfo{pages}{645} (\bibinfo{year}{1998}).

\bibitem[{\citenamefont{Kasprzak et~al.}(2006)\citenamefont{Kasprzak, Richard,
  Kundermann, Baas, Jeambrun, Keeling, Marchetti, Szyma{\'n}ska, Andr{\'e},
  Staehli et~al.}}]{KasprzakNature06}
\bibinfo{author}{\bibfnamefont{J.}~\bibnamefont{Kasprzak}},
  \bibinfo{author}{\bibfnamefont{M.}~\bibnamefont{Richard}},
  \bibinfo{author}{\bibfnamefont{S.}~\bibnamefont{Kundermann}},
  \bibinfo{author}{\bibfnamefont{A.}~\bibnamefont{Baas}},
  \bibinfo{author}{\bibfnamefont{P.}~\bibnamefont{Jeambrun}},
  \bibinfo{author}{\bibfnamefont{J.}~\bibnamefont{Keeling}},
  \bibinfo{author}{\bibfnamefont{F.}~\bibnamefont{Marchetti}},
  \bibinfo{author}{\bibfnamefont{M.~H.} \bibnamefont{Szyma{\'n}ska}},
  \bibinfo{author}{\bibfnamefont{R.}~\bibnamefont{Andr{\'e}}},
  \bibinfo{author}{\bibfnamefont{J.}~\bibnamefont{Staehli}},
  \bibnamefont{et~al.}, \bibinfo{journal}{Nature}
  \textbf{\bibinfo{volume}{443}}, \bibinfo{pages}{409} (\bibinfo{year}{2006}).

\bibitem[{\citenamefont{Lagoudakis et~al.}(2008)\citenamefont{Lagoudakis,
  Wouters, Richard, Baas, Carusotto, André, Dang, and
  Deveaud-Pledran}}]{LagoudakisNatPhys08}
\bibinfo{author}{\bibfnamefont{K.~G.} \bibnamefont{Lagoudakis}},
  \bibinfo{author}{\bibfnamefont{M.}~\bibnamefont{Wouters}},
  \bibinfo{author}{\bibfnamefont{M.}~\bibnamefont{Richard}},
  \bibinfo{author}{\bibfnamefont{A.}~\bibnamefont{Baas}},
  \bibinfo{author}{\bibfnamefont{I.}~\bibnamefont{Carusotto}},
  \bibinfo{author}{\bibfnamefont{R.}~\bibnamefont{André}},
  \bibinfo{author}{\bibfnamefont{D.~L.~S.} \bibnamefont{Dang}},
  \bibnamefont{and}
  \bibinfo{author}{\bibfnamefont{B.}~\bibnamefont{Deveaud-Pledran}},
  \bibinfo{journal}{Nature Physics} \textbf{\bibinfo{volume}{4}},
  \bibinfo{pages}{706} (\bibinfo{year}{2008}).

\bibitem[{\citenamefont{Amo et~al.}(2009)\citenamefont{Amo, Lefrère, Pigeon,
  Adrados, Ciuti, Carusotto, Houdré, Giacobino, and Bramati}}]{AmoNatLett09}
\bibinfo{author}{\bibfnamefont{A.}~\bibnamefont{Amo}},
  \bibinfo{author}{\bibfnamefont{J.}~\bibnamefont{Lefrère}},
  \bibinfo{author}{\bibfnamefont{S.}~\bibnamefont{Pigeon}},
  \bibinfo{author}{\bibfnamefont{C.}~\bibnamefont{Adrados}},
  \bibinfo{author}{\bibfnamefont{C.}~\bibnamefont{Ciuti}},
  \bibinfo{author}{\bibfnamefont{I.}~\bibnamefont{Carusotto}},
  \bibinfo{author}{\bibfnamefont{R.}~\bibnamefont{Houdré}},
  \bibinfo{author}{\bibfnamefont{E.}~\bibnamefont{Giacobino}},
  \bibnamefont{and} \bibinfo{author}{\bibfnamefont{A.}~\bibnamefont{Bramati}},
  \bibinfo{journal}{Nature Physics} \textbf{\bibinfo{volume}{5}},
  \bibinfo{pages}{805} (\bibinfo{year}{2009}).

\bibitem[{\citenamefont{Kaitouni et~al.}(2006)\citenamefont{Kaitouni, Daïf,
  Baas, Richard, Paraiso, Lugan, Guillet, Morier-Genoud, Ganière, Staehli
  et~al.}}]{KaitouniPRB06}
\bibinfo{author}{\bibfnamefont{R.~I.} \bibnamefont{Kaitouni}},
  \bibinfo{author}{\bibfnamefont{O.~E.} \bibnamefont{Daïf}},
  \bibinfo{author}{\bibfnamefont{A.}~\bibnamefont{Baas}},
  \bibinfo{author}{\bibfnamefont{M.}~\bibnamefont{Richard}},
  \bibinfo{author}{\bibfnamefont{T.}~\bibnamefont{Paraiso}},
  \bibinfo{author}{\bibfnamefont{P.}~\bibnamefont{Lugan}},
  \bibinfo{author}{\bibfnamefont{T.}~\bibnamefont{Guillet}},
  \bibinfo{author}{\bibfnamefont{F.}~\bibnamefont{Morier-Genoud}},
  \bibinfo{author}{\bibfnamefont{J.~D.} \bibnamefont{Ganière}},
  \bibinfo{author}{\bibfnamefont{J.~L.} \bibnamefont{Staehli}},
  \bibnamefont{et~al.}, \bibinfo{journal}{Phys. Rev. B}
  \textbf{\bibinfo{volume}{74}}, \bibinfo{pages}{155311}
  (\bibinfo{year}{2006}).

\bibitem[{\citenamefont{Lugan et~al.}(2006)\citenamefont{Lugan, Sarchi, and
  Savona}}]{LuganPSSC06}
\bibinfo{author}{\bibfnamefont{P.}~\bibnamefont{Lugan}},
  \bibinfo{author}{\bibfnamefont{D.}~\bibnamefont{Sarchi}}, \bibnamefont{and}
  \bibinfo{author}{\bibfnamefont{V.}~\bibnamefont{Savona}},
  \bibinfo{journal}{phys. stat. sol. (c)} \textbf{\bibinfo{volume}{3}},
  \bibinfo{pages}{2428} (\bibinfo{year}{2006}).

\bibitem[{\citenamefont{Cerna et~al.}(2009)\citenamefont{Cerna, Sarchi,
  Paraïso, Nardin, L\`eger, Richard, Pietka, Daif, Morier-Genoud, Savona
  et~al.}}]{CernaPRB09}
\bibinfo{author}{\bibfnamefont{R.}~\bibnamefont{Cerna}},
  \bibinfo{author}{\bibfnamefont{D.}~\bibnamefont{Sarchi}},
  \bibinfo{author}{\bibfnamefont{T.~K.} \bibnamefont{Paraïso}},
  \bibinfo{author}{\bibfnamefont{G.}~\bibnamefont{Nardin}},
  \bibinfo{author}{\bibfnamefont{Y.}~\bibnamefont{L\`eger}},
  \bibinfo{author}{\bibfnamefont{M.}~\bibnamefont{Richard}},
  \bibinfo{author}{\bibfnamefont{B.}~\bibnamefont{Pietka}},
  \bibinfo{author}{\bibfnamefont{O.~E.} \bibnamefont{Daif}},
  \bibinfo{author}{\bibfnamefont{F.}~\bibnamefont{Morier-Genoud}},
  \bibinfo{author}{\bibfnamefont{V.}~\bibnamefont{Savona}},
  \bibnamefont{et~al.}, \bibinfo{journal}{Phys. Rev. B}
  \textbf{\bibinfo{volume}{80}}, \bibinfo{pages}{121309(R)}
  (\bibinfo{year}{2009}).

\bibitem[{\citenamefont{Cerna et~al.}(2010)\citenamefont{Cerna, Paraiso, Leger,
  Wouters, Morier-Genoud, Portella-Oberli, and Deveaud-Pl\'edran}}]{CernaPRB10}
\bibinfo{author}{\bibfnamefont{R.}~\bibnamefont{Cerna}},
  \bibinfo{author}{\bibfnamefont{T.}~\bibnamefont{Paraiso}},
  \bibinfo{author}{\bibfnamefont{Y.}~\bibnamefont{Leger}},
  \bibinfo{author}{\bibfnamefont{M.}~\bibnamefont{Wouters}},
  \bibinfo{author}{\bibfnamefont{F.}~\bibnamefont{Morier-Genoud}},
  \bibinfo{author}{\bibfnamefont{M.}~\bibnamefont{Portella-Oberli}},
  \bibnamefont{and}
  \bibinfo{author}{\bibfnamefont{B.}~\bibnamefont{Deveaud-Pl\'edran}},
  \bibinfo{journal}{Phys Rev B} \textbf{\bibinfo{volume}{81}},
  \bibinfo{pages}{113306} (\bibinfo{year}{2010}).

\bibitem[{\citenamefont{Nardin et~al.}(2010)\citenamefont{Nardin, Léger,
  Pietka, Morier-Genoud, and Deveaud-Plédran}}]{NardinPRB10}
\bibinfo{author}{\bibfnamefont{G.}~\bibnamefont{Nardin}},
  \bibinfo{author}{\bibfnamefont{Y.}~\bibnamefont{Léger}},
  \bibinfo{author}{\bibfnamefont{B.}~\bibnamefont{Pietka}},
  \bibinfo{author}{\bibfnamefont{F.}~\bibnamefont{Morier-Genoud}},
  \bibnamefont{and}
  \bibinfo{author}{\bibfnamefont{B.}~\bibnamefont{Deveaud-Plédran}},
  \bibinfo{journal}{Phys. Rev. B} \textbf{\bibinfo{volume}{82}},
  \bibinfo{pages}{045304} (\bibinfo{year}{2010}).

\bibitem[{\citenamefont{Gurioli et~al.}(2001)\citenamefont{Gurioli, Bogani,
  Wiersma, Roussignol, Cassabois, Khitrova, and Gibbs}}]{GurioliPRB01}
\bibinfo{author}{\bibfnamefont{M.}~\bibnamefont{Gurioli}},
  \bibinfo{author}{\bibfnamefont{F.}~\bibnamefont{Bogani}},
  \bibinfo{author}{\bibfnamefont{D.~S.} \bibnamefont{Wiersma}},
  \bibinfo{author}{\bibfnamefont{P.}~\bibnamefont{Roussignol}},
  \bibinfo{author}{\bibfnamefont{G.}~\bibnamefont{Cassabois}},
  \bibinfo{author}{\bibfnamefont{G.}~\bibnamefont{Khitrova}}, \bibnamefont{and}
  \bibinfo{author}{\bibfnamefont{H.}~\bibnamefont{Gibbs}},
  \bibinfo{journal}{Phys. Rev. B} \textbf{\bibinfo{volume}{64}},
  \bibinfo{pages}{165309} (\bibinfo{year}{2001}).

\bibitem[{\citenamefont{Langbein and Hvam}(2002)}]{LangbeinPRL02}
\bibinfo{author}{\bibfnamefont{W.}~\bibnamefont{Langbein}} \bibnamefont{and}
  \bibinfo{author}{\bibfnamefont{J.~M.} \bibnamefont{Hvam}},
  \bibinfo{journal}{Phys. Rev. Lett.} \textbf{\bibinfo{volume}{88}},
  \bibinfo{pages}{047401} (\bibinfo{year}{2002}).

\bibitem[{\citenamefont{Langbein}(2002)}]{LangbeinICPS02}
\bibinfo{author}{\bibfnamefont{W.}~\bibnamefont{Langbein}}, in
  \emph{\bibinfo{booktitle}{Proc. 26th Int. Conf. on the Physics of
  Semiconductors}} (\bibinfo{year}{2002}).

\bibitem[{\citenamefont{Langbein}(2004)}]{LangbeinJPhys04}
\bibinfo{author}{\bibfnamefont{W.}~\bibnamefont{Langbein}},
  \bibinfo{journal}{J. Phys.: Condens. Matter} \textbf{\bibinfo{volume}{16}},
  \bibinfo{pages}{S3645} (\bibinfo{year}{2004}).

\bibitem[{\citenamefont{Langbein}(2010)}]{LangbeinRNC10}
\bibinfo{author}{\bibfnamefont{W.}~\bibnamefont{Langbein}},
  \bibinfo{journal}{Rivista del nuovo cimento} \textbf{\bibinfo{volume}{33}},
  \bibinfo{pages}{255} (\bibinfo{year}{2010}).

\bibitem[{\citenamefont{Herman and Sitter}(1989)}]{Herman89MBE}
\bibinfo{author}{\bibfnamefont{M.}~\bibnamefont{Herman}} \bibnamefont{and}
  \bibinfo{author}{\bibfnamefont{H.}~\bibnamefont{Sitter}},
  \emph{\bibinfo{title}{Molecular Beam Epitaxy}}
  (\bibinfo{publisher}{Springer-Verlag Berlin}, \bibinfo{year}{1989}).

\bibitem[{\citenamefont{Kawada et~al.}(1993)\citenamefont{Kawada, Shirayone,
  and Takahashi}}]{KawadaJCG93}
\bibinfo{author}{\bibfnamefont{H.}~\bibnamefont{Kawada}},
  \bibinfo{author}{\bibfnamefont{S.}~\bibnamefont{Shirayone}},
  \bibnamefont{and}
  \bibinfo{author}{\bibfnamefont{K.}~\bibnamefont{Takahashi}},
  \bibinfo{journal}{Journal of Crystal Growth} \textbf{\bibinfo{volume}{128}},
  \bibinfo{pages}{550 } (\bibinfo{year}{1993}).

\bibitem[{\citenamefont{Oesterle et~al.}(2005)\citenamefont{Oesterle, Stanley,
  and Houdré}}]{OesterlePSSB05}
\bibinfo{author}{\bibfnamefont{U.}~\bibnamefont{Oesterle}},
  \bibinfo{author}{\bibfnamefont{R.~P.} \bibnamefont{Stanley}},
  \bibnamefont{and} \bibinfo{author}{\bibfnamefont{R.}~\bibnamefont{Houdré}},
  \bibinfo{journal}{physica status solidi (b)} \textbf{\bibinfo{volume}{242}},
  \bibinfo{pages}{2157} (\bibinfo{year}{2005}), ISSN \bibinfo{issn}{1521-3951},
  \urlprefix\url{http://dx.doi.org/10.1002/pssb.200560971}.

\bibitem[{\citenamefont{Andrews et~al.}(2002)\citenamefont{Andrews, Specka,
  Romanov, Bobeth, and Pompe}}]{AndrewsJAP02}
\bibinfo{author}{\bibfnamefont{A.~M.} \bibnamefont{Andrews}},
  \bibinfo{author}{\bibfnamefont{J.~S.} \bibnamefont{Specka}},
  \bibinfo{author}{\bibfnamefont{A.~E.} \bibnamefont{Romanov}},
  \bibinfo{author}{\bibfnamefont{M.}~\bibnamefont{Bobeth}}, \bibnamefont{and}
  \bibinfo{author}{\bibfnamefont{W.}~\bibnamefont{Pompe}}, \bibinfo{journal}{J.
  Appl. Phys.} \textbf{\bibinfo{volume}{91}}, \bibinfo{pages}{1933}
  (\bibinfo{year}{2002}).

\bibitem[{\citenamefont{Adachi}(1988)}]{AdachiBook88}
\bibinfo{author}{\bibfnamefont{S.}~\bibnamefont{Adachi}},
  \emph{\bibinfo{title}{Properties of Aluminium Gallium Arsenide}}
  (\bibinfo{publisher}{INSPEC}, \bibinfo{address}{London},
  \bibinfo{year}{1988}).

\bibitem[{\citenamefont{Matthews and Blakeslee}(1974)}]{MatthewsJCG74}
\bibinfo{author}{\bibfnamefont{J.~W.} \bibnamefont{Matthews}} \bibnamefont{and}
  \bibinfo{author}{\bibfnamefont{A.~E.} \bibnamefont{Blakeslee}},
  \bibinfo{journal}{J. Crystal Growth} \textbf{\bibinfo{volume}{27}},
  \bibinfo{pages}{118} (\bibinfo{year}{1974}).

\bibitem[{\citenamefont{Ettenberg and Paff}(1970)}]{EttenbergJAP70}
\bibinfo{author}{\bibfnamefont{M.}~\bibnamefont{Ettenberg}} \bibnamefont{and}
  \bibinfo{author}{\bibfnamefont{R.}~\bibnamefont{Paff}},
  \bibinfo{journal}{Journal of Applied Physics} \textbf{\bibinfo{volume}{41}},
  \bibinfo{pages}{3926} (\bibinfo{year}{1970}).

\bibitem[{\citenamefont{Shinohara et~al.}(1985)\citenamefont{Shinohara, Ito,
  and Imamura}}]{ShinoharaJAP85}
\bibinfo{author}{\bibfnamefont{M.}~\bibnamefont{Shinohara}},
  \bibinfo{author}{\bibfnamefont{T.}~\bibnamefont{Ito}}, \bibnamefont{and}
  \bibinfo{author}{\bibfnamefont{Y.}~\bibnamefont{Imamura}},
  \bibinfo{journal}{J. Appl. Phys} \textbf{\bibinfo{volume}{58}},
  \bibinfo{pages}{3449} (\bibinfo{year}{1985}).

\bibitem[{\citenamefont{Grabmaier and Watson}(1969)}]{GrabmaierPSSB69}
\bibinfo{author}{\bibfnamefont{J.~G.} \bibnamefont{Grabmaier}}
  \bibnamefont{and} \bibinfo{author}{\bibfnamefont{C.~B.}
  \bibnamefont{Watson}}, \bibinfo{journal}{physica status solidi (b)}
  \textbf{\bibinfo{volume}{32}}, \bibinfo{pages}{K13} (\bibinfo{year}{1969}).

\bibitem[{\citenamefont{Bacher and J.~S.~Harris}(1995)}]{BacherJElChemSoc95}
\bibinfo{author}{\bibfnamefont{K.}~\bibnamefont{Bacher}} \bibnamefont{and}
  \bibinfo{author}{\bibfnamefont{J.}~\bibnamefont{J.~S.~Harris}},
  \bibinfo{journal}{J. Electrochem. Soc.} \textbf{\bibinfo{volume}{142}},
  \bibinfo{pages}{2386} (\bibinfo{year}{1995}).

\bibitem[{\citenamefont{Giannuzzi and Stevie}(2005)}]{Giannuzzi05FIB}
\bibinfo{editor}{\bibfnamefont{L.~A.} \bibnamefont{Giannuzzi}}
  \bibnamefont{and} \bibinfo{editor}{\bibfnamefont{F.~A.}
  \bibnamefont{Stevie}}, eds., \emph{\bibinfo{title}{Introduction to focused
  ion beams: instrumentation, theory, techniques, and practise}}
  (\bibinfo{publisher}{Springer-Verlag Berlin}, \bibinfo{year}{2005}).

\end{thebibliography}

\end{document}